\documentclass[aps,prl,twocolumn,superscriptaddress,groupedaddress]{revtex4}

\usepackage{amsmath}
\usepackage{amssymb}
\usepackage{graphicx}
\usepackage{epsfig}
\usepackage{dcolumn}
\usepackage{bm}
\usepackage{bm}
\usepackage{blindtext}
\usepackage{natbib}
\usepackage{booktabs}
\usepackage{mathrsfs}
\usepackage{epstopdf}
\usepackage{color}

\usepackage{textcomp}
\usepackage{times}
\usepackage{float}
\usepackage{latexsym,amsmath,amssymb,bm,euscript}
\usepackage{subfigure}
\usepackage[colorlinks=true,linkcolor=blue,citecolor=blue]{hyperref}
\usepackage{hyperref}
\usepackage{soul}
\usepackage[normalem]{ulem}
\usepackage{lettrine}
\usepackage{xspace}
\usepackage{textcomp}

\def\be{\begin{equation}}
\def\ee{\end{equation}}
\def\bea{\begin{eqnarray}}
\def\eea{\end{eqnarray}}

\begin{document}

\title{$E_8$ Spectra of Quasi-one-dimensional Antiferromagnet BaCo$_2$V$_2$O$_8$ under Transverse Field}

\author{Haiyuan Zou}
\thanks{These authors contributed equally to this study.}
\affiliation{Tsung-Dao Lee Institute, Shanghai Jiao Tong University, Shanghai, 200240, China}

\author{Yi Cui}
\thanks{These authors contributed equally to this study.}
\affiliation{Department of Physics and Beijing Key Laboratory of
Opto-electronic Functional Materials $\&$ Micro-nano Devices, Renmin
University of China, Beijing, 100872, China}

\author{Xiao Wang}
\thanks{These authors contributed equally to this study.}
\affiliation{Tsung-Dao Lee Institute, Shanghai Jiao Tong University, Shanghai, 200240, China}

\author{Z. Zhang}
\affiliation{Tsung-Dao Lee Institute, Shanghai Jiao Tong University, Shanghai, 200240, China}

\author{J. Yang}
\affiliation{Tsung-Dao Lee Institute, Shanghai Jiao Tong University, Shanghai, 200240, China}

\author{G. Xu}
\affiliation{NIST Center for Neutron Research, National Institute of Standards and Technology, Gaithersburg, MD, 20899-6102, USA}

\author{A. Okutani}
\affiliation{Center for Advanced High Magnetic Field Science, Graduate School of Science, Osaka University, Toyonaka, Osaka 560-0043, Japan}

\author{M. Hagiwara}
\affiliation{Center for Advanced High Magnetic Field Science, Graduate School of Science, Osaka University, Toyonaka, Osaka 560-0043, Japan}

\author{M. Matsuda}
\affiliation{Neutron Scattering Division, Oak Ridge National Laboratory, Oak Ridge, Tennessee 37831, USA}

\author{G. Wang}
\affiliation{Beijing National Laboratory for Condensed Matter Physics, Institute of Physics, Chinese Academy of Sciences, Beijing 100190, China}

\author{Giuseppe Mussardo}
\affiliation{SISSA and INFN, Sezione di Trieste, Via Bonomea 265, I-34136 Trieste, Italy}

\author{K. H\'{o}ds\'{a}gi}
\affiliation{BME-MTA Statistical Field Theory Research Group, Institute of Physics, Budapest University of Technology and Economics, 1111 Budapest, Budafoki \'{u}t 8, Hungary}

\author{M. Kormos}
\affiliation{MTA-BME Quantum Dynamics and Correlations Research Group, Department of Theoretical Physics, Budapest University of Technology and Economics, 1111 Budapest, Budafoki \'{u}t 8, Hungary}

\author{Zhangzhen He}
\affiliation{State Key Laboratory of Structural Chemistry, Fujian Institute of Research on the Structure of Matter, Chinese Academy of Sciences, Fuzhou, Fujian 350002, China}

\author{S. Kimura}
\affiliation{Institute for Materials Research, Tohoku University, Sendai, Miyagi 980-8577, Japan}

\author{Rong Yu}
\affiliation{Department of Physics and Beijing Key Laboratory of Opto-electronic Functional Materials $\&$ Micro-nano Devices, Renmin University of China, Beijing, 100872, China}

\author{Weiqiang Yu}
\email{wqyu\_phy@ruc.edu.cn}
\affiliation{Department of Physics and Beijing Key Laboratory of Opto-electronic Functional Materials $\&$ Micro-nano Devices, Renmin University of China, Beijing, 100872, China}

\author{Jie Ma}
\email{jma3@sjtu.edu.cn}
\affiliation{Key Laboratory of Artificial Structures and Quantum Control (Ministry of Education), Shenyang National Laboratory for Materials
Science, School of Physics and Astronomy, Shanghai Jiao Tong University, Shanghai 200240, China}

\author{Jianda Wu}
\email{wujd@sjtu.edu.cn}
\affiliation{Tsung-Dao Lee Institute, Shanghai Jiao Tong University, Shanghai, 200240, China}
\affiliation{School of Physics $\&$ Astronomy, Shanghai Jiao Tong University, Shanghai, 200240, China}


\begin{abstract}

We report $^{51}$V nuclear magnetic resonance (NMR) and inelastic neutron scattering (INS) measurements on a quasi-1D antiferromagnet BaCo$_2$V$_2$O$_8$ under transverse field along the [010] direction. The scaling behavior of the spin-lattice relaxation rate above the N\'{e}el temperatures unveils a 1D quantum critical point (QCP) at $H_c^{1D}\approx 4.7$ T, which is masked by the 3D magnetic order. With the aid of accurate analytical analysis and numerical calculations, we show that the zone center INS spectrum at $H_c^{1D}$ is precisely described by the pattern
of the 1D quantum Ising model in a magnetic field, a class of universality described in terms of the exceptional $E_8$ Lie algebra. These excitations keep to be non-diffusive over a certain field range when the system is away from the 1D QCP. Our results provide an unambiguous experimental realization of the massive $E_8$ phase in the compound, and open new experimental route for exploring the dynamics of quantum integrable systems as well as physics beyond integrability.

\end{abstract}

\maketitle

Strong fluctuations in the vicinity of quantum phase transitions can induce exotic ground states and excitations~\cite{Sachdev_2011}, such as unconventional quantum critical scalings~\cite{Pfeuty_AP_1970,Sachdev_2011}, deconfined quantum critical point (QCP)~\cite{Senthil_Science_2004,Wang_PRX_2017}, and emergent enriched symmetries~\cite{Isakov_PRB_2003}. However, pursuing intrinsic features of these exotic states is a challenging quest and only few exactly solvable models provide significant insight. For example, an exotic spin liquid ground state can be characterized by the honeycomb Kitaev model~\cite{Kitaev_AOP_2006}, and stimulates serious hunting in materials~\cite{Banarjee_Science_2017}. Remarkably, an integrable model~\cite{Zamolodchikov_1989} emerges when the QCP of the paradigmatic 1D transverse-field Ising chain (TFIC)~\cite{Pfeuty_AP_1970,Sachdev_2011} is perturbed by a longitudinal magnetic field. The excitations of this model are beautifully characterized by the interplay of eight particles governed by the $E_8$ exceptional Lie algebra. This $E_8$ picture is a compelling pattern of the general class of the universality of the 1D TFIC once perturbed by a longitudinal magnetic field, as shown originally by Zamolodchikov~\cite{Zamolodchikov_1989}. Therefore, finding and exploring the $E_8$ physics in condensed matter systems will be a significant milestone for realizing analytically predicted emergent exotic excitations and provides a manipulable platform to explore quantum magnetism.

Compelling though it may be, the manifestation of this exotic $E_8$ physics can only be established via a dynamics study. In experiments, it is very challenging to accurately determine the location of a 1D QCP and resolve all the eight massive states perturbed away from the 1D QCP. Despite these inherent difficulties, the observation of the lowest two $E_8$ states ($m_1$ and $m_2$) at the 1D ferromagnetic QCP of the quasi-1D magnet CoNb$_2$O$_6$ from inelastic neutron scattering (INS) measurements provided evidence of the quantum $E_8$ spectrum~\cite{Coldea_Science_2010}, and motivated further materials-based studies on this fascinating phenomenon~\cite{Kjall_PRB_2011}. Recently, the quasi-1D Heisenberg-Ising antiferromagnetic (AFM) materials, e.g., BaCo$_2$V$_2$O$_8$ (BCVO) and SrCo$_2$V$_2$O$_8$ (SCVO), has attracted numerous studies with rich quantum phases and excitations induced by transverse or longitudinal field ~\cite{Faure_NP_2018,Kimura_JPSJ_2013,Cui_PRL_2019,Zou_JP_2020,
Ideta_PRB_2012,Niesen_PRB_2013,Bouillot_PRB_2011,Horvatic_PRB_2020}. It is desirable to explore whether this system can host a complete picture of $E_8$ physics.
Along this line, excitations up to the fifth $E_8$ particle ($m_5$) has been resolved by a recent terahertz (THz) spectroscopy measurement on BCVO~\cite{Wang_Prb_2020}.

\begin{figure}[t]
\includegraphics[width=8.5cm]{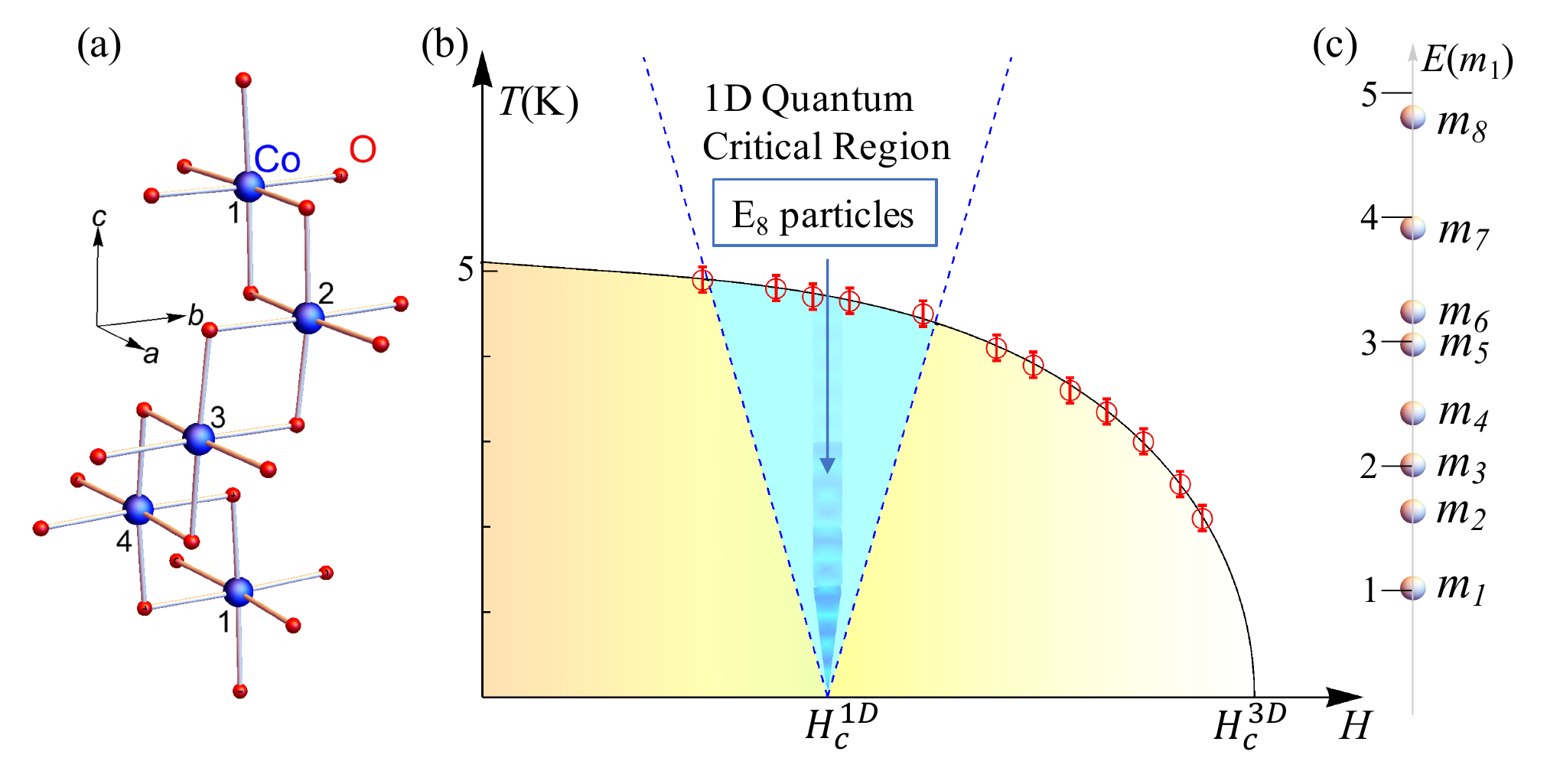}
\caption{\label{fig1} {\bf BaCo$_2$V$_2$O$_8$ in a transverse magnetic field.} (a) The crystal structure of BaCo$_2$V$_2$O$_8$.
Co (O) is labelled by blue (red) spheres. Note that the apical Co-O bond are tilted $\sim 5 ^\circ$ from the c-axis~\cite{Wichmann_1986}. (b) Schematic phase diagram of
BCVO in transverse field. Brown circles represent $T_N$ determined by $1/T_1$ (see Fig.~\ref{fig2}(a,b)), with a 3D QCP at $H_{\rm C}^{3D}$~=~10.4$\pm$0.1~T and a putative 1D QCP at $H_{\rm C}^{1D}$~=~4.7$\pm$0.3~T. Blue ribbon area covers the location of emergent exotic $E_8$ phase.
(c) An illustration of $E_8$ single particle masses $m_i$ ($i$ = 1, 2,$...$, 8) along the energy axis (see also Tab. S1 in~\cite{SM}). The digits label the energy
in unit of $m_1$.}
\end{figure}

In this letter, we report unambiguous identification of the full $E_8$ spectrum via a combination of nuclear magnetic resonance (NMR) and INS measurements on BCVO with field along the [010] direction.
We use NMR to accurately locate the 1D QCP [$H^{1D}_c$ in Fig.~\ref{fig1}(b)], then perform the INS measurements to present the full $E_8$ spectrum for the first time. This result is highly consistent with both the numerical calculations with Eq.~(\ref{eq1}) for the BCVO material and the theoretical analysis of the essential integrable part of the model, providing an unambiguous evidence for the existence of the $E_8$ physics. Furthermore, our study also captures all the multi-particle modes in the studied energy window. This rare experimental realization of the $E_8$ physics and other coherent modes suggested by an integrable system provides a solid experimental test bed for exploring exotic feature of the dynamics and the excitations in quantum magnets.

To begin with, BCVO can be described by the Hamiltonian~\cite{Kimura_JPSJ_2013}
\begin{eqnarray}
\nonumber
H&=&J\sum_{n,i}[S^z_{n,i}S^z_{n+1,i}+\epsilon(S^x_{n,i}S^x_{n+1,i}+S^y_{n,i}S^y_{n+1,i})]\\
&+&J'\sum_{n,i\neq j}S^z_{n,i}S^z_{n,j}-\mu_{B}\sum_{n,i}\tilde{g}\mathbf{H}\cdot\mathbf{S}_{n,i},
\label{eq1}
\end{eqnarray}
which includes intra-chain coupling $J$, the anisotropic factor $\epsilon$, and the weak inter-chain coupling $J'$, with the spin-1/2 operator $S^\mu_{n,i}$ ($n$ and and $i/j$ are chain and site labels respectively) and the Land{\'e} factor tensor $\tilde{g}$. Detailed parameter values are described in the Supplemental Material (SM)~\cite{SM}.
Without the $J'$ term, the system reduces to decoupled 1D AFM chains which accommodate a QCP [$H^{1D}_c$ in Fig.~\ref{fig1}(b)] of TFIC universality~\cite{Wang_PRL_2018,Cui_PRL_2019,Zou_JP_2020}. Although the existence of $J'$ hides the putative 1D QCP deeply inside the AFM ordered phase, the renaissance of strong 1D quantum fluctuation provides a finite temperature quantum critical region which can be detected outside the AFM phase through NMR experiments. On the other hand, at the hidden 1D QCP of a TFIC universality basis, the weak $J'$ interaction provides a longitudinal background perfectly satisfying all the prerequisites to exhibit exotic $E_8$ excitations~\cite{SM}, with the eight single-particle masses in unit of $m_1$ are shown in Fig.~\ref{fig1}(c).
Note that due to the screw structure [Fig.~\ref{fig1}(a)], characterized by the $g$-tensor $\tilde{g}$~\cite{Kimura_JPSJ_2013}, external field $\mathbf{H}$ along the $a$-axis induces an effective staggered in-plane field to lower both $H^{1D}_c$ and $H^{3D}_c$ significantly.

\begin{figure}[t]
\includegraphics[width=8.5cm]{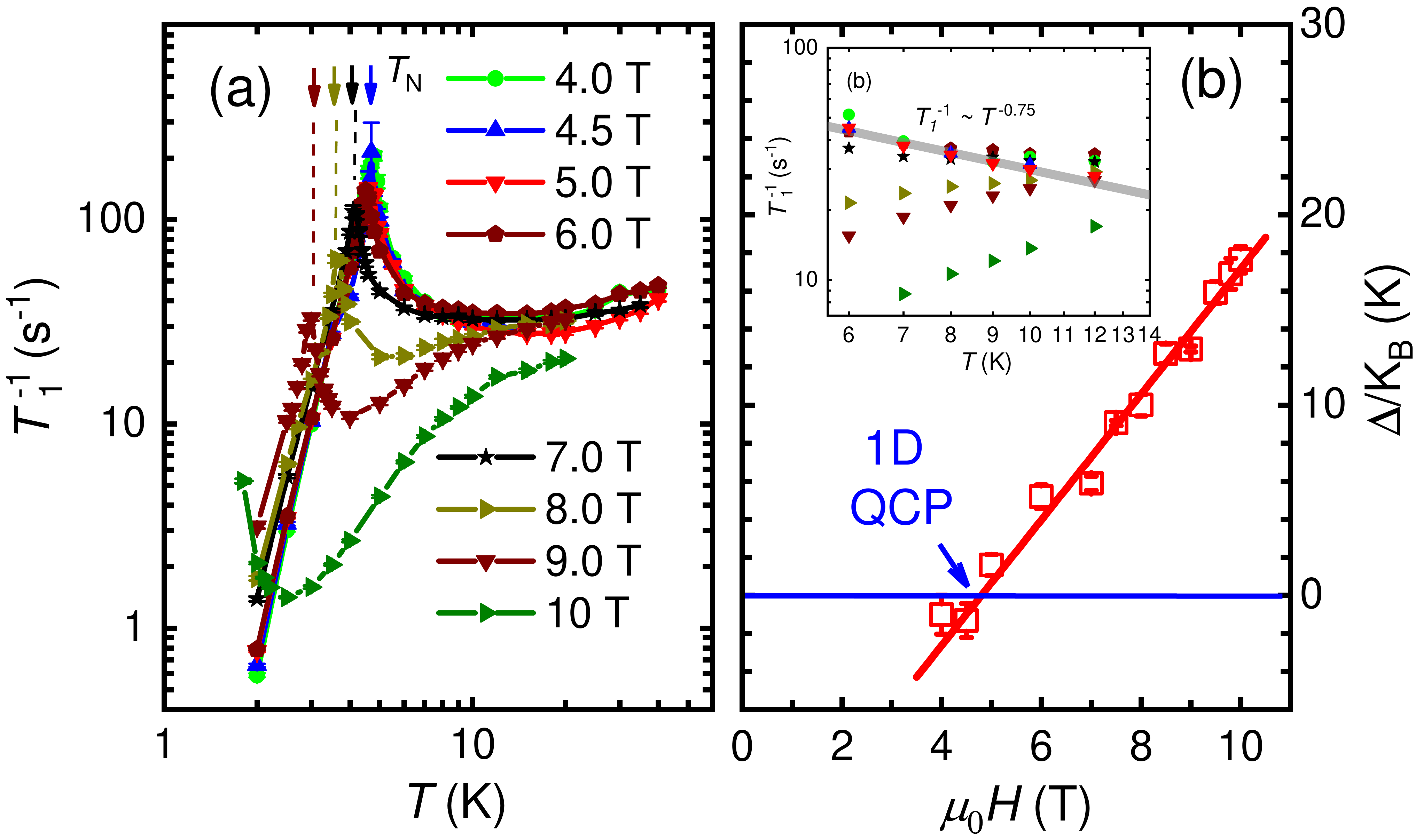} 
\caption{\label{fig2} {\bf 1D QCP of TFIC universality identified by NMR measurements with field along the [010] direction.} (a) The spin-lattice relaxation rate $1/T_1$ as functions of temperatures measured under different in-plane fields. Down arrows point at the peaked position in $1/T_1$, which determine the $T_N$. (b) The 1D gap $\Delta$ obtained by fitting $1/T_1$ (see text) with temperature from 6 K to 12 K. The solid line is a linear fit to $\Delta (H)$, which gives the gap closing field of $4.7\pm0.3$~T as the 1D QCP. Inset: An enlarged view of the data in the low-temperature, paramagnetic regime in the log-log scale, with a
straight guide line $1/T_1  \sim T^{-0.75}$.
}
\end{figure}

\begin{figure*}
\includegraphics[width=14cm]{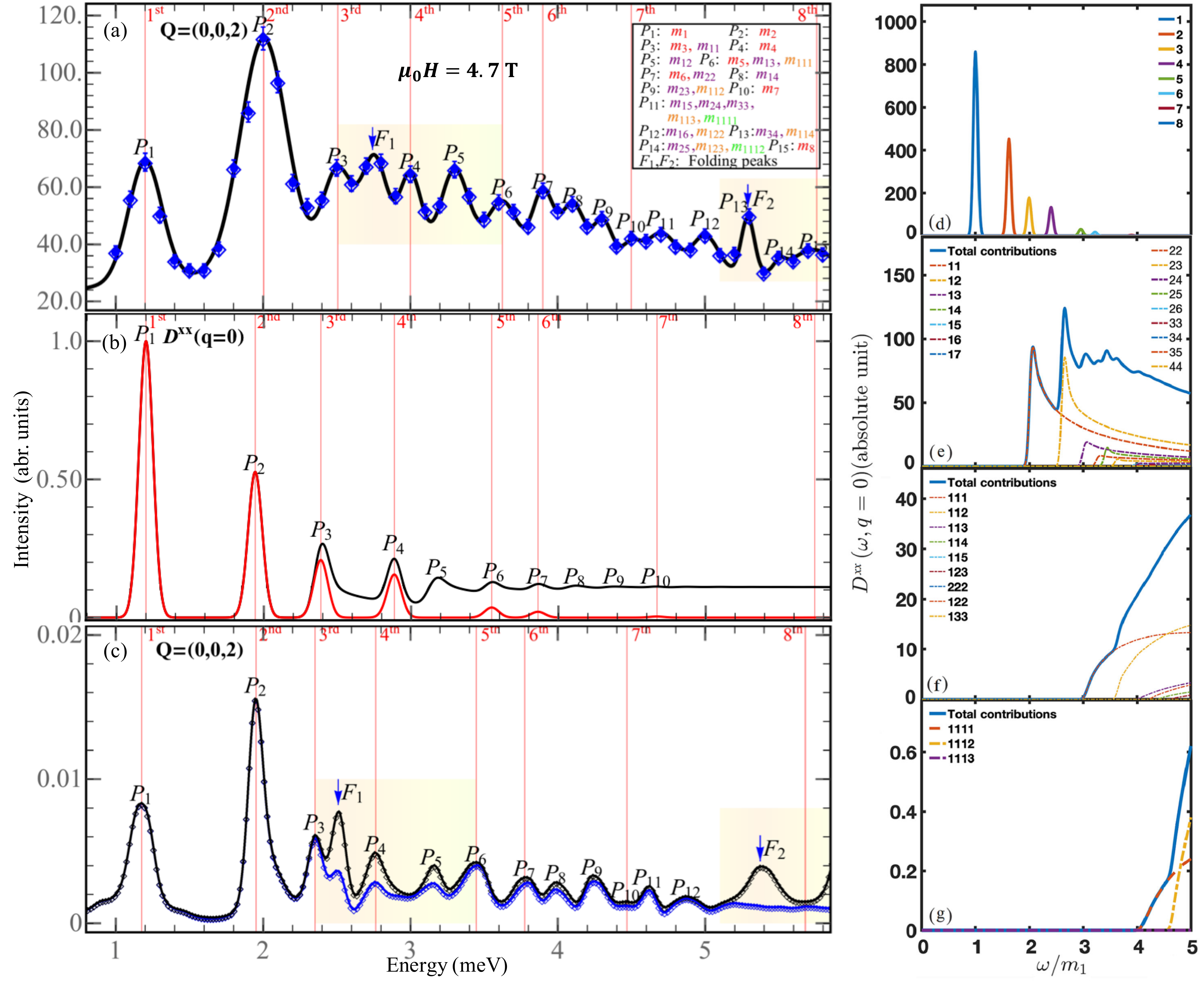} 
\caption{\label{fig3} {\bf $E_8$ excitation spectrum near the 1D QCP.} (a) (Zone center) INS Intensity at Q = (002) in BCVO at the vicinity of the 1D QCP with $H$ = 4.7 T at 0.4 K. Blue diamonds with error bars correspond to experimental data and black lines are the fit to the Gaussian functions. The red vertical lines at eight peaks correspond to the eight single $E_8$ particles. Other peaks come from multi-particle excitation and zone-folding effect identified by iTEBD method. The peak with mass $m_{i_1 i_2...i_n }$ labels multi-particle channel with particles masses $m_{i_1} m_{i_2}...m_{i_n}$. The colored regions illustrate contributions from zone-folding effect with transfer momentum at $q$=0 for regions near the F$_1$ peak and at $q=\pi/2$ for regions near the F$_2$ peak. (b) The analytical dynamic structure factor $D^{xx}$ calculated from quantum $E_8$ integrable field theory~\cite{SM}. Red curve stands for single-particle spectra, while black curve is obtained after including multi-particle contributions. In accord with the experiment, $m_1$ is set as 1.2 meV and the analytical data are broadened in a Lorentzian fashion with full-width at half-maximum fixed at $0.08m_1$.
(c) Neutron scattering intensity from iTEBD calculations at the zone center. The black and blue curves are results with and without zone-folding effect. DSF spectra for individual scattering channel: (d) single-, (e) two-, (f) three-, and (g) four-particles contributions to the $D^{xx}$. $ijkl$ refer to excitations with combined mass modes of $m_im_jm_km_l$.}
\end{figure*}

We first determined the location of the putative 1D QCP via carrying out $^{51}$V NMR measurements on BCVO with transverse field scanning from 3~T to 12~T. The spin-lattice relaxation rate 1/$T_1$ with temperature at various transverse field values are shown in Fig.~\ref{fig2}(a). Below 15~K, 1/$T_1$ exhibits a strong field dependence consistent with the onset of 1D critical magnetic correlations. At low fields, the magnetic transitions are clearly evidenced by a peaked feature in the 1/$T_1$, with strong low-energy spin fluctuations when magnetic ordering occurs. A sharp drop of 1/$T_1$ is followed below N\'{e}el temperature $T_N$, dominated by the spin-wave excitations. The $T_N$ at each field is then determined from the drop and displayed in the phase diagram of Fig.~\ref{fig1}(b), consistent with results from other measurement, e.g., magnetic susceptibility~\cite{Kimura_JPSJ_2013}.

In the vicinity of the 1D QCP, the 1/$T_1$ of the paramagnetic phase at high-temperature follows an analytical form of $1/T_1 \sim T^{-0.75} e^{-\Delta/K_BT}$, where $\Delta$ is the gap value~\cite{Kinross_PRX_2014,Sachdev_2011}.
From 6~K to 12~K, $1/T_1$ follows the power-law form at 5~T and deviates from it at other fields, hence, a 1D QCP ($H_{\rm C}^{\rm 1D}$)
is suggested at about 5~T [see Fig.~\ref{fig2}(b) inset].
To better resolve the $H_{\rm C}^{\rm 1D}$, we fit the gap $\Delta$ and find it follows $\Delta (H)=1.62g_{xx} \mu_B (H-H_{\rm C}^{\rm 1D})$ with field, where $H_{\rm C}^{\rm 1D}=4.7\pm0.3$ T (Fig.~\ref{fig2}(b)). The linearly vanishing gap and the scaling exponent 0.75 near 5 T demonstrate a 1D QCP of TFIC universality in the relaxation rate~\cite{Sachdev_2011}.
As demonstrated in the inset of Fig.~\ref{fig2}(b), $1/T_1\sim  T^{-0.75}$ is at field near 5~T and and the spin dynamics at the 1D QCP is characterized.

Since the 1D QCP hides in the 3D ordering dome,
the competition between this 1D quantum criticality and the ordered background results in interesting excitation. We then performed INS to measure the dynamical structure factor (DSF)
of BCVO under the transverse magnetic field. Figure~\ref{fig3}(a) shows many distinctive peaks in the constant-Q (zero transfer momentum) spectra at zone center Q = (002), with field near the identified (masked) 1D QCP. Assisted by two theoretical calculations: (i) analytical calculation from Eq.~(\ref{eq2}); (ii) infinite time-evolving block decimation (iTEBD)~\cite{Vidal_PRL_2007,White_77_2008} numerical calculation based on Eq.~(\ref{eq1}), these are found to match quantitatively a combination of three types of excitations, including single-$E_8$ states, multi-$E_8$ states, and zone-folding modes as labeled by $P_1$ to $P_{15}$ in the figure. Details are described below.

 Starting from the Zeeman ladder of confinement bound states~\cite{Coldea_Science_2010} at zero magnetic field, the analysis to separate the spin-flip and non-spin-flip contributions at finite magnetic field give a qualitative description of the low energy excitations~\cite{Faure_NP_2018}. However, a full picture of the full excitation feature especially at the high energy region is still missing. Fortunately,
at $H \approx H_{\rm C}^{\rm 1D}$, the stable N\'{e}el ordering masks the desired QCP of TFIC on one hand, but provides a necessary effective perturbation field to bring in the $E_8$ physics on the other hand~\cite{SM}. The slightly off-critical TFIC can be described by a central charge $c=1/2$ conformal field theory (CFT) with the perturbation of its relevant magnetic field~\cite{Zamolodchikov_1989,Delfino_NPB_1995,Delfino_PLB_1996},
\begin{equation}
H_{E_8}=H_{1/2}+h\int\sigma (x)dx
\label{eq2}
\end{equation}
where $H_{1/2}$ is the Hamiltonian of the $c=1/2$ CFT which describes the $J,\mu_B$ term in Eq.~(\ref{eq1})~\cite{Cui_PRL_2019,Zou_JP_2020}, and the perturbation from the field $\sigma(x)$ with coupling $h$ corresponds to the effective field of the chain mean-field of the interchain term in Eq.~(\ref{eq1}) when the material is in the AFM phase, and is absent outside the ordering phase.

Following the form factor framework~\cite{Delfino_NPB_1995,Delfino_PLB_1996,Mussardo_2010,Delfino_JPA_2004,Delfino_NPB_1996}, we calculate the DSF, $D^{\alpha\alpha} (\omega,q=0)$ ($\alpha=x,y,z$ along the crystalline [100], [010] and [001] direction, respectively), for the zero-momentum transfer. The analytical result for $D^{xx}$ is displayed in Fig.~\ref{fig3}(b) for the total spectral weight and the $E_8$ particles for series of peaks are clarified in Fig.~\ref{fig3}(d).
Astonishingly, analytical calculations also identify multi-particle channels, which exhibit clearly distinguishable peaks for the spectrum continuum above the two-particle threshold, as shown in Fig.~\ref{fig3}(e)-(g).

We then directly compare these excitations with
the neutron dynamic spectra (Fig.~\ref{fig3}(a)), and find the later
matches excellently with the analytical prediction for the peak positions and the spectra weights (Fig.~\ref{fig3}(b)).

First, all the eight single-$E_8$ particle energies, whose positions are marked by red vertical lines, are resolvable, and a common trend of reducing spectral weight with increasing energies. Note that the two lightest particles matching the golden mass ratio~1.618 as expected~\cite{Zamolodchikov_1989,Delfino_NPB_1995,Delfino_PLB_1996}.

Second, various multi-$E_8$ particle excitations are also clearly resolved as pronounced spectral peaks, even in the high-energy region,
leading to the accountability of these modes and full consistency with our theoretical results (Table S1~\cite{SM}). Due to the low dimensionality, each two-particle scattering channel contributes a non-trivial two-particle DSF continuum with a sharp edge at low-energy boundary and a peaked feature close to this boundary (Fig.~\ref{fig3}(e)), leading to experimentally distinguishable peaks.

Finally, some additional features in the measured spectra are caused by microscopic details of the system, 
which can be fully taken care of by the iTEBD calculation.
For example, due to the four-period screw structure of the BCVO, the DSF obtained from iTEBD [Fig.~\ref{fig3}(c)] is able to distinguish a zone folding peak in between the $m_3$ and $m_4$ particle excitation, and another one coincident with multi-$E_8$-particle excitation in between the $m_7$ and $m_8$ particle excitation. As another example, the strong suppression of the $m_1$ 
observed in neutron data at Q = (002) is also captured by the detailed iTEBD calculations (Fig.~\ref{fig3}). As shown in section S3 of the SM~\cite{SM}, the suppression of the $E_8$ particle excitations is caused by the 
spin-flip term 
(the $\epsilon$ term) of Eq.~\ref{eq1}, and this effect is the most significant for 
the $m_1$ peak. 

\begin{figure}[t]
\includegraphics[width=8cm]{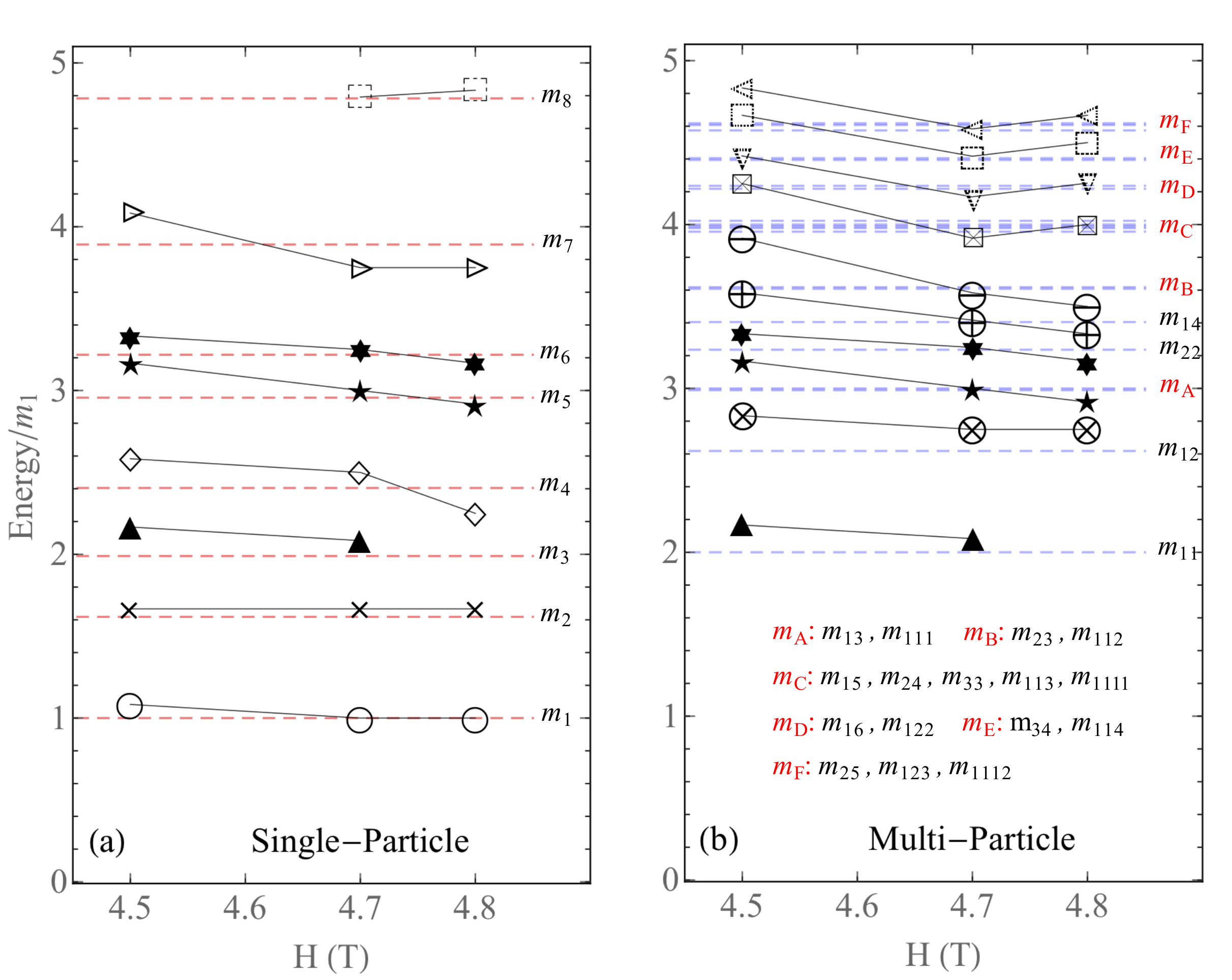} 
\caption{\label{fig4} (a) and (b) INS spectra peaks at different transverse field, with field-evolutions of the single- and multi-particle masses respectively, from Q = (002) in BCVO with $H$=4.5~T, 4.7~T, and 4.8~T at 0.4~K. The dashed shapes in Figs. (a, b) illustrate the $E_8$ particles above 4$m_1$, which are not completely experimental resolvable due to the energy resolution limit (0.1~meV) and the zone-folding effect.}
\end{figure}

To show that the realized $E_8$ physics in BCVO close to the QCP is not accidental, we further carry out neutron experiments with other two nearby magnetic fields $H$~=~4.5~T and 4.8~T at Q = (002). The frequency of the spectral peaks for different fields are collected and displayed in Fig.~\ref{fig4}, where an evolution of mass distribution continuously passes through the $E_8$ phase. Not surprisingly,
the best agreement with the $E_8$ predictions is found at the field of 4.7~T, at the putative 1D QCP predetermined by NMR. This field is lower than 5~T obtained by the Terahertz studies of the $E_8$ modes~\cite{Wang_Prb_2020}. Details of dynamical spectra for those fields can be found in the SM~\cite{SM}.

The excellent agreement between the analytical dynamical $E_8$ spectrum and the neutron data at the putative 1D QCP in BCVO supports the unambiguous existence of complex $E_8$ physics.
One striking feature revealed in our INS data is the robustness of the high-energy, single mass peaks, which do not decay into the multiple low-mass modes at the critical field. The underlying physics needs to be further addressed theoretically. Moreover, when the field deviates from the $E_8$ phase, the $E_8$ peaks gradually smear out ({\it cf}. Fig.~S4~\cite{SM}). However, shift of spectra peaks and spectral weight transfer with clear DSF peaks are still resolved away from the critical field upon the decay of $E_8$ particles. This shows that quantum many-body ground state continuously deforms with the tuning of external parameters (field, pressure, etc.), meanwhile, the dynamic spectrum rearranges itself into new non-diffusive modes.
Studying non-integrable effects is of great importance to give a comprehensive understanding on real materials since integrability is rare in reality.
The BCVO now can serve as an ideal test bed:
the corresponding 1D effective model turns into a perturbed quantum $E_8$ model, which is no longer integrable with heavy $E_8$ particles beginning to decay~\cite{Delfino_NPB_1996,Delfino_NPB_2006}.

In conclusion, the combined experiments on BCVO, detection of 1D QCP via NMR and the observation of dynamical spectrum through neutron scattering, together with their excellent agreements with the analytical results from the quantum $E_8$ integrable field theory, unambiguously realize the beautiful $E_8$ physics in this AFM material, leaving us with an exemplary realization of the $E_8$ spectrum. The precise iTEBD numerical simulation on the microscopic model further resolves the details of the spectrum and bridges accurately the essential physics in the quantum integrable model and the realistic dynamics in the material. Our study sets a concrete ground to explore the physics of $E_8$ particles and excitations, as well as physics beyond the integrable model.
A better control and identification of the excitations above the ground state, and the understanding of the robustness of the high-energy modes, are therefore crucial to arrive at a comprehensive understanding for the quantum many-body system and to study not only its equilibrium properties but also its non-equilibrium features (for instance, Ref.~\onlinecite{Calabrese_2016} and references therein).

We thank G\'{a}bor Tak\'{a}cs for useful discussions about the form factor calculations. J.W. thanks helpful discussion with Zhe Wang. H.Z. is supported by the National Natural Science Foundation of China with Grant No.~11804221.  W.Q.Y and R.Y. are supported by the Ministry of Science and Technology of China with Grant No.~2016YFA0300504, the National Natural Science Foundation of China with Grants Nos.~51872328, 11674392, and the Fundamental Research Funds for the Central Universities, and the Research Funds of Renmin University of China with Grant Nos. 18XNLG24 and 20XNLG19. J.W. is sponsored by Natural Science Foundation of Shanghai with Grant No. 20ZR1428400 and Shanghai Pujiang Program with Grant No. 20PJ1408100. J.M. is supported by the Ministry of Science and Technology of China with Grant No. 2016YFA0300501, the National Natural Science Foundation of China with Grants No. 11774223. J.W. and J.M. acknowledges additional support from a Shanghai talent program. Y.C. is supported by China Postdoctoral Science Foundation with Grant No. 2020M680797, the Fundamental Research Funds for the Central Universities, and the Research Funds of Renmin University of China with Grant No. 21XNLG18. G.W. is supported by the National Natural Science Foundation of China with Grant No.~51832010 and the Ministry of Science and Technology of China with Grant No. 2018YFE0202600. K.H. and M.K. were partially supported by the National Research Development and Innovation Office of Hungary under the research grants OTKA
K-16 No. 119204 and by the Fund TKP2020 IES (Grant No. BME-IE-NAT). M.K. acknowledges support by a Bolyai J\'{a}nos grant of the HAS, and by the \'{U}NKP-20-5 New National Excellence Program of the Ministry for Innovation and Technology from the source of the National Research Development and Innovation Fund.

H.Z., Y.C. and X.W. contributed equally to this study. J.W. conceived the project. J.M. and W.Y. conducted the experiments.

{\it Note added:} A review article that highlights the current work appears online by Journal Club for Condensed Matter Physics~\cite{Oshikawa_JCCMP_2020}.

\end{document}